\documentclass{aa}
\usepackage{graphicx}
\begin{document}

\title{Effects of Thermal Fluctuations in the SPOrt Experiment}
\author{E.~Carretti\inst{1} \and
        M.~Zannoni\inst{2} \and
        C.~Macculi\inst{1} \and
        S.~Cortiglioni\inst{1} \and
        C.~Sbarra\inst{1}}

\institute{I.A.S.F./C.N.R. Bologna, Via Gobetti 101, I-40129 Bologna, Italy.
           \and
           I.A.S.F./C.N.R. Milano, Via Bassini 15, 
	         I-20133 Milano, Italy.
          }

\offprints{Ettore Carretti, \email{carretti@bo.iasf.cnr.it}}

\abstract{The role of systematic errors induced by 
thermal fluctuations is analyzed for the SPOrt experiment
with the aim at estimating their impact
on the measurement of the 
Cosmic Microwave Background Polarization (CMBP).
The transfer functions of the antenna devices
from temperature to data fluctuations are computed, by writing them
in terms of both instrument and thermal environment parameters.
In addition, the corresponding contamination maps are estimated,
along with their
polarized power spectra, for different behaviours of the
instabilities.
The result is that thermal effects are at a negligible level even for fluctuations 
correlated with the Sun illumination
provided their frequency $f_{\rm tf}$ is larger than that of the
Sun illumination ($f_{\rm day}$) by a factor 
$f_{\rm tf} / f_{\rm day} > 30$, which defines
a requirement for the statistical properties of the temperature
behaviour as well.
The analysis with actual SPOrt operative parameters shows that
the instrument is only weakly sensitive 
to temperature instabilities, the main contribution 
coming from the cryogenic stage.
The contamination on the $E$-mode spectrum does not
significantly pollute the CMBP signal and no specific 
data cleaning seems to be needed.
}

\date{Received / Accepted}

\titlerunning{Effects of Thermal Fluctuations in the SPOrt Experiment}
\authorrunning{E. Carretti et al.}
\maketitle

\section{Introduction}

The tiny level of the Cosmic Microwave Background Polarization (CMBP)
requires a careful understanding and estimate of 
all systematic effects, which, if not kept under control,
can jeopardize the measurement of the signal.

In the last years several works have been published
on this topic, witnessing an increasing interest
in systematics (e.g. Piat et al. 2000, Carretti et al. 2001,
Leahy et al. 2002, Kaplan \& Delabrouille 2002, 
Mennella et al. 2002, 
Hu et al. 2003,  Page et al. 2003, 
Franco et al. 2003, Carretti et al. 2004).
Among others, thermal fluctuations can seriously affect the data
and many teams have studied the impact on their CMB experiments 
(e.g. Piat et al. 2000, Mennella et al. 2002, Page et al. 2003). 
The importance of thermal fluctuations 
depends on the receiver scheme and,
in some cases, can be crucial. 
An example is the 
total power architecture, where the variations induced
onto the data are only slightly dumped with respect to the 
primary temperature fluctuations, making the detection
of a $\mu$K signal a real challenge 
even in a very stable thermal environment.
Thus, besides a quiet environment, receivers with a 
low sensitivity to temperature
are necessary to weaken thermal fluctuations effects.  
Having a low offset generation, correlators 
minimize the thermal disturbance.

In this work we present an estimate of the thermal contamination
for the SPOrt\footnote{http://sport.bo.iasf.cnr.it} 
experiment (Cortiglioni et al. 2004), 
a set of four correlation radio-polarimeters devoted
to measure the Stokes parameters $Q$ and $U$ of CMBP
with an angular resolution of FWHM~$=7^\circ$
from the International Space
Station (ISS). The ISS environment is not optimal
from the thermal point of view. In fact, thermal 
simulations show that
the Sun illumination modulation due to the
motion around the Earth induces  
orbit--synchronous temperature fluctuations with $\sim 3$~K amplitude 
in case no active temperature control is used. 
Thus, a careful analysis of thermal effects is mandatory.

In the following we derive the transfer functions
from temperature to data fluctuations to evaluate
the sensitivity of the
SPOrt radiometers to thermal instabilities. In addition,
we study the contamination on the $E$-mode signal as a
function of the statistical properties of the 
temperature instabilities, aiming at estimating its
impact on the cosmological signal.

We show that the SPOrt correlation receivers 
are naturally an optimal architecture to keep
under control the noise induced by temperature variations,
thanks to their low sensitivity to temperature.
In fact, the contamination
on the $E$-mode is close to the cosmological signal 
already in case of free temperature fluctuations
without thermal control, even though not sufficiently low
to allow a clean detection.
The adoption of an active
control for the horn section  further reduces
the spurious signal at a comfortably negligible level,
leaving the CMBP signal uncontaminated.

The paper is organized as follows: the computation of the
transfer functions is presented in Sec.~\ref{offSec}, 
whereas the effects of the fluctuation
statistics and the SPOrt scanning strategy are presented in Sec.~\ref{statSec}.
The transfer functions and the contamination on the $E$-mode specific 
for the SPOrt experiment are given in
Sec.~\ref{sportSec} and, finally, the conclusions are
summarized in Sec.~\ref{concSec}.

\section{Transfer Functions}\label{offSec}

Thermal fluctuation effects can be evaluated
starting from the offset generation equation.
In SPOrt-like correlation polarimeters the offset
is mainly generated by horn, polarizer and orthomode transducer (OMT), 
according to the equation\footnote{$j$ denotes the complex unit}
(Carretti et al. 2001)
\begin{eqnarray}
 Q+jU  
             &=& S\!P_{\rm omt}
               \left(T_{\rm sky} + T_{\rm atm} +
                 T_{\rm noise}^{\rm Ant}
               \right)  \nonumber \\
             &+&   S\!P_{\rm pol}
                  \left(T_{\rm sky} + T_{\rm atm} +
                            T_{\rm noise}^{\rm h} -
                            A_{\rm h}\;T_{\rm ph}^{\rm p}
                          \right),\nonumber \\
                  \label{AB0TN0eq}
\end{eqnarray}
where $S\!P_{\rm omt}$ and $S\!P_{\rm pol}$ are
\begin{eqnarray}
 S\!P_{\rm omt} & = & A_{\rm omt}\;\left(S_{A1}S_{B1}^* + S_{A2}S_{B2}^*\right),
                     \label{spomtEq} \\
                     \nonumber \\
 S\!P_{\rm pol} & = & {1\over 2} \left(1 - {A_{\parallel}\over
                                 A_{\perp}}\right) 
                  =  {1\over 2} \; {A_{\perp} - A_{\parallel}
                       \over A_{\perp}},
\label{sppolEq}
\end{eqnarray}
$S_{A1}$, $S_{B2}$ the transmission parameters of the two OMT arms,
$S_{A2}$, $S_{B1}$ their isolation terms, $A_{\rm h}$, $A_{\rm omt}$
the attenuations of the horn and the OMT, respectively, 
and $A_{\perp}$, $A_{\parallel}$ the attenuations 
of the polarizer along its two main polarizations.
The offset sources are thus
the physical temperature of the polarizer 
$T_{\rm ph}^{\rm p}$ and the signals propagating in the antenna system:
the signal collected from the sky ($T_{\rm sky}$, in antenna temperature), the 
atmosphere emission ($T_{\rm atm}$) and the noise temperatures
of the horn alone ($T_{\rm noise}^{\rm h}$) 
and of the whole antenna system
\begin{equation}
  T_{\rm noise}^{\rm Ant} = T_{\rm noise}^{\rm h} +
  A_{\rm h}\; T_{\rm noise}^{\rm p} + 
  A_{\rm h}\;A_{\rm p}\; T_{\rm noise}^{\rm omt},
  \label{tantnoiseeq}
\end{equation}
with $A_{\rm p}$ the mean attenuation of the two polarizer arms.
A complete description of derivation
and implications of Eq.~(\ref{AB0TN0eq}) is given in
Carretti et al. (2001). Here we 
just point out that the offset is generated by OMT and polarizer
that partially correlate the antenna noise
as well as the sky and atmosphere emissions. 

Practically, all the terms are sensitive to thermal fluctuations:
$T_{\rm ph}^{\rm p}$ by
definition; $T_{\rm noise}^{\rm h}$, $T_{\rm noise}^{\rm p}$ and
$T_{\rm noise}^{\rm omt}$ according to the equations 
(e.g. see Kraus 1986)
\begin{eqnarray} 
T_{\rm noise}^{\rm h}  & = &                           
               (A_{\rm h} - 1)\;T_{\rm ph}^{\rm h}\label{thneq}\\
T_{\rm noise}^{\rm p}   & = &   
               (A_{\rm p } - 1)\;T_{\rm ph}^{\rm p}\label{tpneq}\\
T_{\rm noise}^{\rm omt}   & = &
               (A_{\rm omt } - 1)\;T_{\rm ph}^{\rm omt}\label{toneq}
\end{eqnarray}
with $T_{\rm ph}^{\rm h}$, $T_{\rm ph}^{\rm p}$, $T_{\rm ph}^{\rm omt}$
the physical temperatures of horn, polarizer and OMT. Finally, 
also attenuations $A$ and coefficients $S\!P_{\rm pol}$ and 
$S\!P_{\rm omt}$ are functions of
the temperature through the relations 
(see Appendix~\ref{AppA} for a detailed derivation)
\begin{eqnarray}
 \Delta A     & = & {A-1 \over 2\;T}\;\Delta T \nonumber\\
 \Delta(A-1)   & = & {A-1 \over 2\;T}\;\Delta T \nonumber\\
 \Delta(S\!P_{\rm pol})   & = & {S\!P_{\rm pol} \over 2\;T_{\rm ph}^{\rm p}}
                                \;\Delta T_{\rm ph}^{\rm p} \nonumber\\
 \Delta(S\!P_{\rm omt})   & = & {A_{\rm omt}-1 \over 4\;T_{\rm ph}^{\rm omt}}
                                \;S\!P_{\rm omt}
                                \;\Delta T_{\rm ph}^{\rm omt}
		  \label{attFlucteq}
\end{eqnarray}
Using the Eqs.~(\ref{AB0TN0eq})--(\ref{attFlucteq}) we can write the
offset variations as a function of those of the 
physical temperatures as
\begin{eqnarray}
 \Delta(Q+jU)   & = &
           H_{\rm h}\; \Delta T_{\rm ph}^{\rm h}\nonumber \\
			 &+& H_{\rm p}\;  \Delta T_{\rm ph}^{\rm p}\nonumber  \\
			 &+& H_{\rm omt}\;  \Delta T_{\rm ph}^{\rm omt}\nonumber  \\
		  \label{offFluct1eq}
\end{eqnarray}
where the transfer functions $H$ of the three devices are defined by
\begin{eqnarray}
          H_{\rm h}   &=& {3\over 2}\,(A_{\rm h} - 1)
     \;[S\!P_{\rm omt}(1+h^{\rm omt}) + S\!P_{\rm pol}(1+h^{\rm p})],\nonumber\\
          \nonumber\\
          H_{\rm p}   &=& {3\over 2}\,A_{\rm h}\;[S\!P_{\rm omt}\,(A_{\rm p} - 1)\, 
                                     (1+p^{\rm omt})-
			                      S\!P_{\rm pol}(1+p^{\rm p})],		\nonumber \\
          \nonumber\\
			    H_{\rm omt} &=& {3\over 2}\,A_{\rm h}\;A_{\rm p}\;(A_{\rm omt} - 1)
			                     \;S\!P_{\rm omt}\;(1+O^{\rm omt}),
          \nonumber\\
			   \label{hhheq}
\end{eqnarray}
with the corrective terms $h$, $p$, $O$ given by
\begin{eqnarray}
          \nonumber\\
          h^{\rm omt}  &=& (A_{\rm p} - 1)\,{T_{\rm ph}^{\rm p}\over
                            3\;T_{\rm ph}^{\rm h}}
                          +A_{\rm p}\,(A_{\rm omt} - 1)\,{T_{\rm ph}^{\rm omt}\over
                            3\;T_{\rm ph}^{\rm h}},\nonumber\\
          h^{\rm p}  &=& -{T_{\rm ph}^{\rm p},\over
                            3\;T_{\rm ph}^{\rm h}},\nonumber\\
          p^{\rm omt}  &=& (A_{\rm omt}-1)\,{T_{\rm ph}^{\rm omt}\over
                            3\;T_{\rm ph}^{\rm p}},\nonumber\\
          p^{\rm p}  &=& -\left(
                            {A_{\rm h}-1 \over A_{\rm h}}\, {T_{\rm ph}^{\rm h}\over
                             3\;T_{\rm ph}^{\rm p}}
                            + {1 \over A_{\rm h}}\, {T_{\rm sky}+T_{\rm atm}\over
                             3\;T_{\rm ph}^{\rm p}}
                          \right), \nonumber\\
          O^{\rm omt}  &=& {A_{\rm h}-1 \over A_{\rm h}\,A_{\rm p}}\, 
                               {T_{\rm ph}^{\rm h}\over 6\;T_{\rm ph}^{\rm omt}}
                        +  {A_{\rm p}-1 \over A_{\rm p}} \,
                               {T_{\rm ph}^{\rm p}\over 6\;T_{\rm ph}^{\rm omt}}
                        +  {A_{\rm omt}-1 \over 6} \nonumber\\
                       & &+  {1 \over A_{\rm h}\,A_{\rm p}}\, 
                               {T_{\rm sky}+T_{\rm atm}\over 6\;T_{\rm ph}^{\rm omt}}.
                               \nonumber\\
 			   \label{hpOeq}
\end{eqnarray}
These corrective terms, dominated by $(A-1)$ terms, 
are in general much lower than 1.
Slightly different is $h^{\rm p}$, which can be as high as
$|h^{\rm p}| = 1/3$ when $T_{\rm ph}^{\rm h} =
T_{\rm ph}^{\rm p}$, but which, anyway, reduces the instability effects.

The dumping factors with respect to the thermal fluctuations
are thus given by noise generation terms ($A-1$) and by
the extra-terms ($S\!P_{\rm omt}$, $S\!P_{\rm pol}$) typical of the correlation 
architecture. Actually,
the offsets for total power outputs are directly given by 
$T^{\rm Ant}_{\rm noise}$ (Eq.~(\ref{tantnoiseeq}))
and the corresponding transfer functions are  
\begin{eqnarray}
          H_{\rm h}^{TP} &=& {3\over 2}\;(A_{\rm h} - 1)\;(1+h^{\rm omt}),\\
          H_{\rm p}^{TP}  &=& {3\over 2}\;A_{\rm h}\;(A_{\rm p} - 1)\;(1+p^{\rm omt}),		 \\
			    H_{\rm omt}^{TP}&=& {3\over 2}\;A_{\rm h}\;A_{\rm p}\;(A_{\rm omt} - 1),
			   \label{hhhTPeq}
\end{eqnarray}
where the extra terms are not present.
Considering that $S\!P_{\rm omt}$ and $S\!P_{\rm pol}$ can be as low as 
$10^{-3}$ (see Carretti et al. 2001 for an evaluation of the SPOrt case)
the advantages of correlation architectures become obvious.

\section{Effects of Fluctuation Statistics and SPOrt Scanning Strategy}\label{statSec}

The transfer functions provide the instantaneous response
to thermal fluctuations. The relevant effects, however,
have to be evaluated on the final maps, which means also
the scanning strategy and
the behaviour of the thermal fluctuations must be
accounted for.

First, we will estimate the dumping factors by an
approximate analytical analysis. 
It is a worst case analysis which does not
provide a complete description of the effects -- as how the
contamination is spread on the various angular scales --
but it will help us have an insight into the mechanisms dumping the 
thermal contaminations.

Then, we will face the exact treatment by simulations which will
provide us with the real contamination maps and
allow us to estimate the contamination on the signal
power spectra. 

In this section, we will consider
unit offset fluctuations only (arbitrary units),
to better evaluate the pure effects irrespective of the
real offsets generated by the SPOrt receivers.

\subsection{Contamination in one Precession Period}

The scanning strategy of SPOrt consists in observing 
toward the Zenith of the International Space Station
while orbiting in $t_{\rm orbit} = 90$~min 
around the Earth along a $51.6^\circ$ inclination orbit.
The latter is a circle precessing in $t_{\rm pr} = 70$~days, so that 
the observation of the sky within declinations 
$|\delta| < 51.6^\circ$ is performed in the same time
(see Cortiglioni et al. 2004 for details).

The precession moves the trajectory by the $7^{\circ}$
of the beam-size
along the Celestial Equator in $N_{\rm orbit}$~$\sim$~$22$~orbits,
during which SPOrt observes the same pixel stripe.

The map-making procedure consists in averaging
all the data collected in a pixel (see Sbarra et al. 2003 for details).
Thus the error 
$\Delta (Q+jU)_{\rm scan}$ on the maps after
$N_{\rm orbit}$ observations is given by
\begin{equation}
     \Delta (Q+jU)_{\rm scan} = {1\over N_{\rm orbit}} \sum_{i} \Delta (Q+jU)(t_i),
     \label{dQUscaneq}
\end{equation}
where $t_i$ is the time at the $i^{\rm th}$ of the $N_{\rm orbit}$ passages.

The result depends on the behaviour of the fluctuations.
With no loss of generality, we can consider sinusoidal
variations for the temperature of each device
\begin{equation}
  \Delta T = \Delta T_0 \; \cos(2\pi f_{\rm tf}t), 
\end{equation}
since a generic function can always be expanded 
by Fourier transform ($f_{\rm tf}$ is
the fluctuation frequency).
The linear relationship between temperature and offset variations
(Eq.~(\ref{offFluct1eq})) leads to the same behaviour of
the offset fluctuations
\begin{equation}
  \Delta (Q + jU) = \Delta (Q + jU)_0 \; \cos(2\pi f_{\rm tf}t).
  \label{QUsinEq}
\end{equation}
From a statistical point of view, fluctuations 
can be divided in three different types:
\begin{enumerate}
	\item Thermal fluctuations synchronous with the Sun illumination 
	      (i.e. the ISS day). In this case the fluctuation 
	      frequency $f_{\rm tf}$ is that of the ISS day
  	 \begin{eqnarray}
	    f_{\rm tf} & = & f_{\rm day} \nonumber\\
	               & = & f_{\rm orbit} + f_{\rm pr} - f_{\odot},
     \end{eqnarray}
     i.e. the orbit frequency $f_{\rm orbit} = 1/t_{\rm orbit}$
     properly corrected for the ISS precession ($f_{\rm pr}$) 
     and the Sun revolution ($f_{\sun}$). 
	   This is the natural behaviour when no active temperature control
	   is in operation, with
	   the maximum expected under Sun illumination and the minimum
	   when in the Earth shadow.
     In this case no dumping over the $N_{\rm orbit}$ orbits 
     occurs since, when the
     instrument is observing toward the
     same sky pixel during these consecutive orbits, 
     the fluctuation has the same phase and the averaging does not
     dump the effects on the pixel map (see Mennella et al. 2002).
     In this case dumping is simply by a unit factor
     \begin{equation}
        D^1_{\rm scan} = 1.
     \end{equation}     
	\item Thermal fluctuations synchronous with the Sun illumination, but with a 
	   frequency which is an integer multiple of the ISS day
   	 \begin{equation}
	     f_{\rm tf} = K\,f_{\rm day},\;\;\;\;\;\;\;\; {\rm with}\;K\;{\rm integer.}
	   \end{equation}
	   This situation can occur when an active control is in operation, but
	   without enough strength to decorrelate the period with respect to
	   the Sun illumination.
     In this case again no dumping occurs over the $N_{\rm orbit}$ orbits, and
     the corresponding dumping factor is
     \begin{equation}
        D^K_{\rm scan} = 1.
     \end{equation}     
	\item Thermal fluctuations asynchronous with the night-day cycle
	   \begin{equation}
	      f_{\rm tf} \ne K\,f_{\rm day}.
	   \end{equation}
	   This situation is typical of cryogenic sections with closed
	   loop control, where the stabilized temperature 
	   has a behaviour imposed by
	   the control electronics and in general is independent
	   of the outer temperature fluctuations.
	   In this case the situation is better: as described by Mennella et al. (2002)
	   there is a dumping by a factor 
     \begin{equation}
    	   D^{\rm asy}_{\rm scan} = {1 / N_{\rm orbit}},
     \end{equation}     
	   which greatly reduces the effects on the final maps.
     We would like to add only that, differing from the others, 
     the value of this dumping factor is a conservative estimate,
     since it represents the
     envelope of actual $K$ values
     (see Mennella et al. 2002 for a detailed discussion).
\end{enumerate}
Table~\ref{scanDumpArbTab} reports the worst case 
fluctuations on the final maps 
after a precession time (in arbitrary units
relative to the amplitude in a single orbit) 
     \begin{equation}
    	   \Delta(Q +jU)_{\rm scan} = D_{\rm scan}^x\;\Delta(Q +jU) 
     \end{equation}
with $x=K$ or $x={\rm asy}$, 
and shows the advantage to be asynchronous with the Sun illumination.
\begin{table}
 \centering
  \caption{Fluctuations on the map after 1 precession time
           for $7^\circ$ pixels.
           Values for both synchronous and asynchronous 
           thermal instabilities are reported (see text for details).}
  \begin{tabular}{@{}lccc@{}}
     \hline
     $f_{\rm tf}$                     & $N_{\rm orbit}$ & $\Delta (Q + jU)_0$ & 
     $\Delta (Q + jU)_{\rm scan}$  \\
             &    &  [1 orbit] &  [$N_{\rm orbit}$ orbits] \\
     \hline
     $f_{\rm tf} = f_{\rm day}$      & 22     &  1  & 1     \\
     $f_{\rm tf} = K\;f_{\rm day}$   & 22     &  1  & 1     \\
     $f_{\rm tf} \ne K\;f_{\rm day}$ & 22     &  1  & 0.046 \\
     \hline\\
\end{tabular}
\label{scanDumpArbTab}
\end{table}

\begin{figure}
\resizebox{1.0\hsize}{!}{\includegraphics[angle=90]{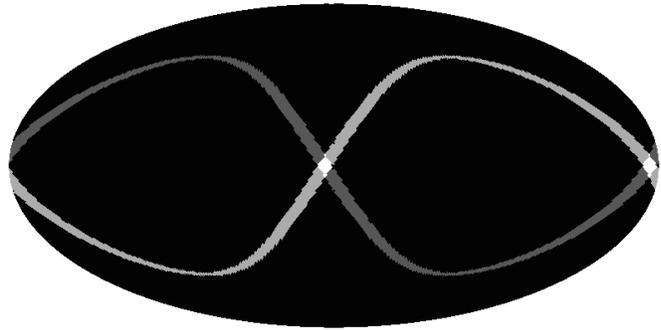}}
\caption{Sky coverage of two orbits  
separated by 35~days (dark and light gray). 
The 180$^\circ$-shift is due to the precession.
The central pixel is observed during both orbits.
The map is a mollview projection in Celestial coordinates.
}
\label{twoOrbFig}
\end{figure}

Actually, in one precession each pixel
is visited twice: once during the first 35~days and a
second time after about half a precession
(see Fig.~\ref{twoOrbFig}).
Since the combination of these two sets of observations does not
play a relevant role in our worst case analysis,
we will discuss it at the end of Sec.~\ref{contLTSec}.

\subsection{Contamination in the Lifetime}\label{contLTSec}

So far we have analyzed dumping after just one sky coverage.
As described above, SPOrt observes the whole of the accessible
sky every 70~days and the final map comes
from the combination of these coverages.
The thermal effects on the final map thus 
depend on how the contamination maps of each
coverage combine with each other.

In the simple case $f_{\rm tf} = f_{\rm day}$ the
fluctuations depend on the Sun illumination and show the maximum temperature
when in presence of the Sun and the minimum when in the Earth shadow. 
Over a precession period, SPOrt
observes each pixel for $N_{\rm orbit}$ consecutive orbits. 
70~Earth-days later,
when the precession allows SPOrt to observe the same pixel again, 
the illumination
conditions are different due to the new position
of the Sun with respect to
the Earth. Thus, each pixel will be observed when  
temperature fluctuations are in different phases 
depending on the observing 
season, which provides some cancellation effects.
As an example, if a pixel is observed during the day (Sun above the
horizon) with a positive temperature fluctuation, 6~months later it 
is observed at night with a negative temperature variation.

In fact, as in Eq.~(\ref{dQUscaneq}), 
the error on the final map is the average of the offset
variations in each observation time:
\begin{eqnarray}
     \Delta (Q+jU)_{\rm fm} & = & {1\over N_{\rm pr}} \sum_{i} \Delta (Q+jU)(t_i),
     \label{dQUfmeq}
\end{eqnarray}
where  $N_{\rm pr}$ is the number of precessions 
and $t_j$ the observing time.
Two different observations of the same pixel differ for a multiple
of both the orbit time $t_{\rm orbit}$ and the precession time $t_{\rm pr}$.
Thus, the $i^{\rm th}$ observation of a pixel happens at the time 
\begin{eqnarray}
     t_i & = & t_{\rm px} + \Delta t_i \nonumber\\	
         & = & t_{\rm px} + n_i \, t_{\rm orbit} \nonumber\\
         & = & t_{\rm px} + i \, t_{\rm pr},
	 \label{tjeq}
\end{eqnarray}
where $i$ is the $i^{\rm th}$ precession, $n_i$ the number of orbits at this
precession and $t_{\rm px}$ the time when the first observation of the pixel
occurred.
Using these expressions for valid $t_i$, Eq.~(\ref{dQUfmeq}) becomes
\begin{eqnarray}
     \Delta (Q+jU)_{\rm fm} & = & D^1_{\rm pr}\;\Delta (Q+jU)_0 \nonumber\\
     D^1_{\rm pr} & = & {\cos\alpha \over N_{\rm pr}} 
                                \sum_{i=0}^{N_{\rm pr}-1} \cos\beta_i +
		  {\sin\alpha \over N_{\rm pr}} 
                                \sum_{i=0}^{N_{\rm pr}-1} \sin\beta_i 
				                             \nonumber\\
     \alpha                & = & 2\pi f_{\rm day} t_{\rm px}\nonumber\\
     \beta_i               & = & 2\pi i {f_{\odot}\over f_{\rm pr}}.
     \label{dQUfm2eq}			  
\end{eqnarray}
$D^1_{\rm pr}$ is the dumping factor related to the precession
with respect to the instantaneous offset amplitude $\Delta (Q+jU)_0$. 
It depends not only on the orbital parameters, but also on the pixel
position (by $t_{\rm px}$).
In order to evaluate the effects we consider the worst case, i.e. the maximum
value of $D^1_{\rm pr}$. Table~\ref{precDprTab} reports its value 
for different mission lifetimes.

\begin{table}
 \centering
  \caption{Dumping factors due to the precession of the SPOrt orbit.
  The cases of thermal behaviour
  synchronous with the Sun illumination with $K=1$ 
  ($D^1_{\rm pr}$), $K=10$ 
  ($D^{10}_{\rm pr}$) and $K=30$ 
  ($D^{30}_{\rm pr}$) are displayed. The asynchronous case 
  ($D^{\rm asy}_{\rm pr}$) is shown as well.}
  \begin{tabular}{@{}cccccc@{}}
     \hline
     Lifetime    & $N_{\rm pr}$ & $D^1_{\rm pr}$ & 
                          $D^{10}_{\rm pr}$ & $D^{30}_{\rm pr}$ 
                          &  $D^{\rm asy}_{\rm pr}$ \\ 
     $[$months$]$    & & & & &  \\ 
     \hline 
     12 &  5  & 0.046  & 0.745 & 0.201 &   0.200\\
     18 &  8  & 0.220  & 0.416 & 0.002 &   0.125\\
     24 &  10 & 0.046  & 0.191 & 0.141 &   0.100\\
     30 &  13 & 0.135  & 0.079 & 0.078 &   0.077\\
     36 &  15 & 0.045  & 0.183 & 0.065 &   0.067\\
     \hline
\end{tabular}
\label{precDprTab}
\end{table}
It is worth noting that the best situation corresponds
to 12, 24 and 36 months, during which
the mission benefits from complete Sun revolutions.
To understand this, one has to bear in mind that 
the phase of the fluctuation at a given pixel
depends on the season and performs
a whole $2\pi$ cycle in one year. When the lifetime is
close to a complete Sun revolution (12, 24, 36 months), 
the phases taken at the various precessions sample well
the whole cycle and the net effect is a cancellation of
the thermal contributions.
On the other hand, when the lifetime is 18 or 30 months, 
the last half of a year covers just a half of the whole cycle
and the cancellation is not optimal.

Moving to the second kind of fluctuations --
$f_{\rm tf}$~=~$K\,f_{\rm day}$ -- similar considerations
can be done and the errors on the final maps 
$\Delta (Q+jU)_{\rm fm}^K$ are given by:
\begin{eqnarray}
     \Delta (Q+jU)_{\rm fm}^K & = & D^K_{\rm pr}\;\Delta (Q+jU)_0 \nonumber\\
     D^K_{\rm pr} & = & {\cos\alpha^K \over N_{\rm pr}} 
                                \sum_{i=0}^{N_{\rm pr}-1} \cos\beta^K_i \nonumber \\
                  & + & {\sin\alpha^K \over N_{\rm pr}} 
                                \sum_{i=0}^{N_{\rm pr}-1} \sin\beta^K_i 
				                             \nonumber\\
     \alpha^K                & = & 2\pi K f_{\rm day} t_{\rm px}\nonumber\\
     \beta^K_i               & = & 2\pi Ki {f_{\odot}\over f_{\rm pr}}.
     \label{dQUfmKeq}			  
\end{eqnarray}
The worst case is again the maximum value of the dumping factor
$D^K_{\rm pr}$ and is reported in Tab.~\ref{precDprTab}
for different lifetimes and a couple
of $K$ values (namely, $K = 10, 30$). While for $K=1$ the phase 
does a complete $2\pi$ period in one Sun revolution,
for $K>1$ the cycle is closed in $1/K$~years and, for each pixel, 
the phase sampling may not be complete and well
distributed over the whole period.
Thus, the $K>1$ case is less predictable and in some pixels
(Tab.~\ref{precDprTab} is only the worst case in the whole map) 
the fluctuations may be only slightly
dumped, as occurs for $K=10$ and lifetime~$=12$~months. 
On the other hand, in a few cases
(as for $K=30$, lifetime~$=18$~months)
even the worst
pixel shows a large dumping.
Anyway, in general, the maximum fluctuations
are worse than in the simplest $K=1$ case.

Finally, in the case of asynchronous fluctuations the same
situation as the $N_{\rm orbit}$ consecutive asynchronous scans occurs,
resulting in a dumping
factor by the number of precessions $N_{\rm pr}$
\begin{equation}
   D^{\rm asy}_{\rm pr} = 1/N_{\rm pr}.
\label{dfASeq}
\end{equation}
Results for various lifetimes are again reported in Tab.~\ref{precDprTab}.

The total reduction on the final map includes the actions
of both the first $N_{\rm orbit}$ scans and the precession, and
is given by the product
\begin{equation}
   D^x = D^x_{\rm scan}\;D^x_{\rm pr}, 
\end{equation}
with $x = K$ or $x =$~asy.
It represents the total efficiency with which the 
scanning strategy is able to dump the signal
fluctuations generated by the system.
The corresponding offset fluctuations
\begin{equation}
  \Delta(Q+jU)_{\rm fm} = D^x\; \Delta(Q+jU)_0
\end{equation}
are reported in Tab.~\ref{dumpFMTab} for two different lifetimes.

\begin{table}
 \centering
  \caption{Fluctuations on the final map due to both the first $N_{\rm orbit}$
           scans and the precession. The values are reported for the four
           cases described in Tab.~\ref{precDprTab} 
           treated in the text and for two different lifetimes.
           All the fluctuations are in arbitrary units considering
           $\left|\Delta(Q+jU)_0\right|=1$ the maximum offset 
           fluctuation in one orbit.}
  \begin{tabular}{@{}lcc@{}}
     \hline
     $f_{\rm tf}$                    &  $\Delta (Q + jU)_{\rm fm}$ 
                                 &  $\Delta (Q + jU)_{\rm fm}$  \\
                                 &  [12 months] &  [36 months] \\
     \hline
     $f_{\rm day}$      &  0.046  & 0.045 \\
     $10\;f_{\rm day}$  &  0.745  & 0.183 \\
     $30\;f_{\rm day}$  &  0.201  & 0.065 \\
     $f_{\rm tf} \ne K\;f_{\rm day}$ &  0.009  & 0.003 \\
     \hline\\
\end{tabular}
\label{dumpFMTab}
\end{table}

However, these values represent the residual pixel fluctuation only for 
the worst pixel. Accounting
for neither the statistical nor angular distribution of the deviations, 
they do not provide a complete description 
of the impact of thermal fluctuations.

As a final consideration, we recall that in one precession the pixels
are in general visited twice:
once during the first 35~days and a second time after about
half a precession (the exact delay depends on the distance of the
pixel from the Equator).
These two sets of observations can be treated as observations of 
two different pixels and
combined only at the end to evaluate the total dumping factor. 
They are characterized by two different parallactic
angles and their combination 
can lead to some cancellation. In fact, 
when the difference between the parallactic angles
is 0$^\circ$ no cancellation occurs;
when the difference is $90^\circ$, the offset is
cancelled out. For the SPOrt scanning strategy, this difference runs
between $0^\circ$ (at the top and bottom edges of the orbit) 
and $103.2^\circ$, i.e. twice the orbit inclination (at the Equator).
Thus, in our worst case analysis no further cancellation
has to be considered.

Anyway, a complete treatment of the SPOrt case must be done via simulations, 
that, reproducing the real scanning strategy and map-making, 
account for all the details.

\subsection{Contamination on Angular Power Spectra}

A complete description of thermal fluctuation effects is
provided by the angular power spectra of the contamination maps.
To estimate these power spectra we simulate  
the experiment assuming the offset fluctuations of
Eq.~(\ref{QUsinEq}) and generating
the final $Q$, $U$ maps using the SPOrt map-making procedure (Sbarra et al. 2003),
which also accounts for pixel observations at different parallactic angles.
We adopt a unit fluctuation
amplitude $\left|\Delta(Q+jU)_0\right| = 1$ 
(arbitrary units) to evaluate relative effects.

\begin{figure}
\resizebox{1.0\hsize}{!}{\includegraphics{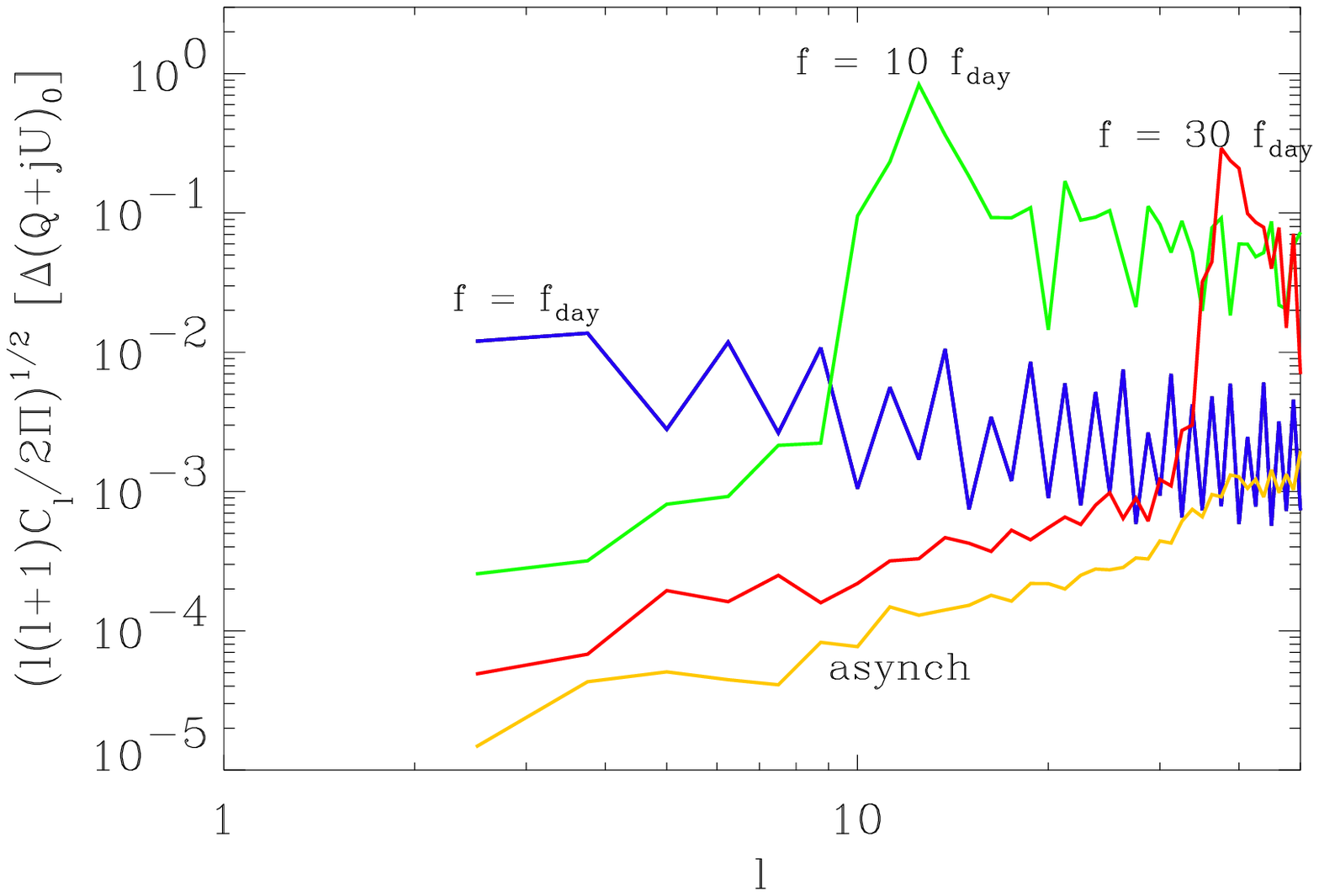}}
\resizebox{1.0\hsize}{!}{\includegraphics{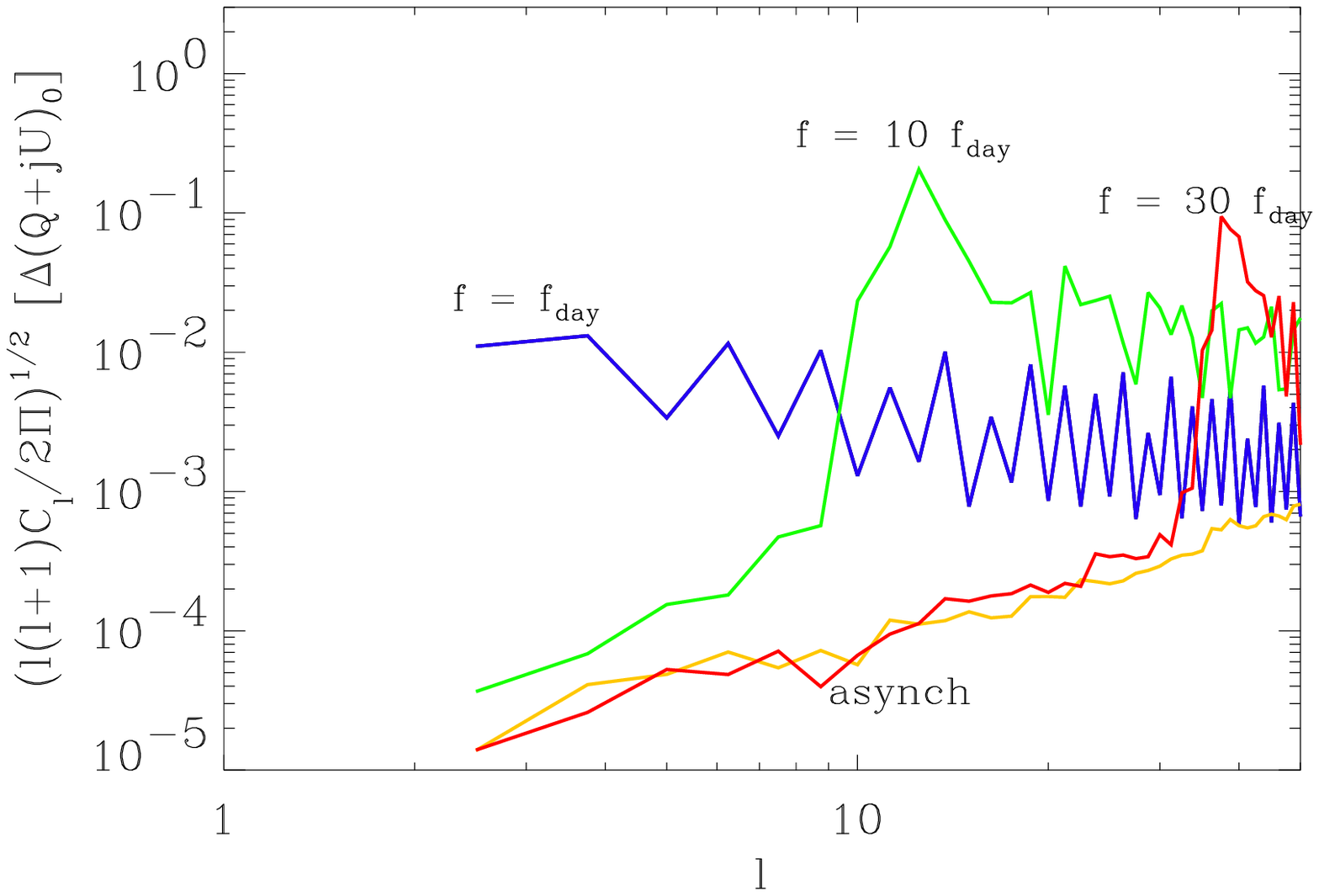}}
\caption{Top: $E$--mode power spectra
of the contamination of the thermal fluctuations
featuring a unit amplitude $|\Delta(Q+jU)_0| = 1$. A lifetime
of 12~months is assumed.
The spectra are presented for four different configurations: synchronous
with the Sun illumination ($f_{\rm tf} = K\;f_{\rm day}$ with $K = 1, 10, 30$),
and asynchronous. For synchronous cases
we plot the average of realizations with different polarization angles of
$\Delta(Q+jU)_0$.
The asynchronous case, instead, is the average of many realizations 
with $K$ in the [30.2,~30.8] range centred on 30.5. 
Ranges with other central frequencies provide similar results. 
Bottom: the same but for a 36-month lifetime.}
\label{spec12Fig}
\end{figure}

Figure~\ref{spec12Fig} reports the results
for a 12-month lifetime.
\begin{figure*}
\centering
  \resizebox{0.6\hsize}{!}{\includegraphics[width=0.5\hsize,angle=90]{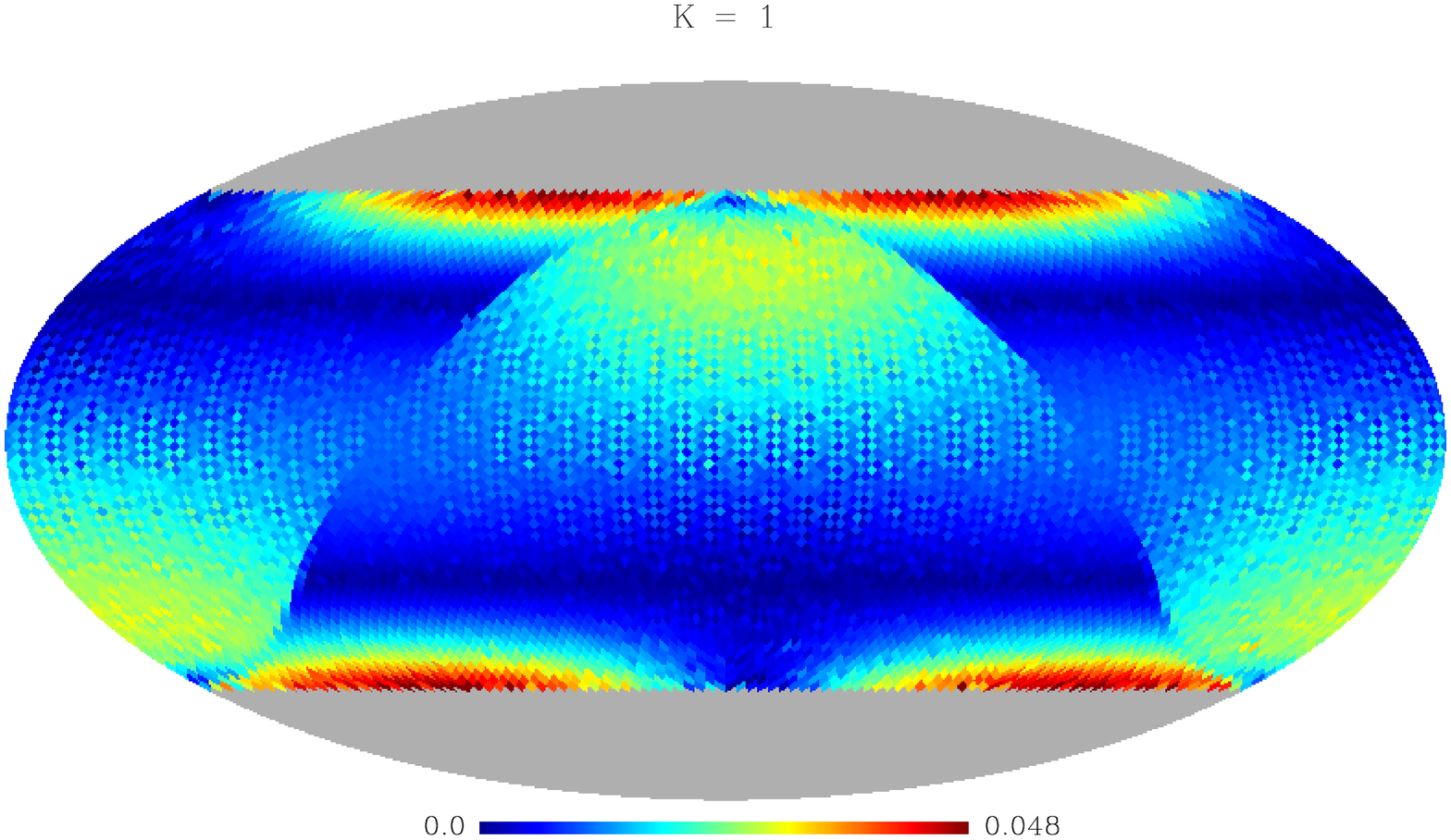}}\\
  \resizebox{0.6\hsize}{!}{\includegraphics[width=0.5\hsize,angle=90]{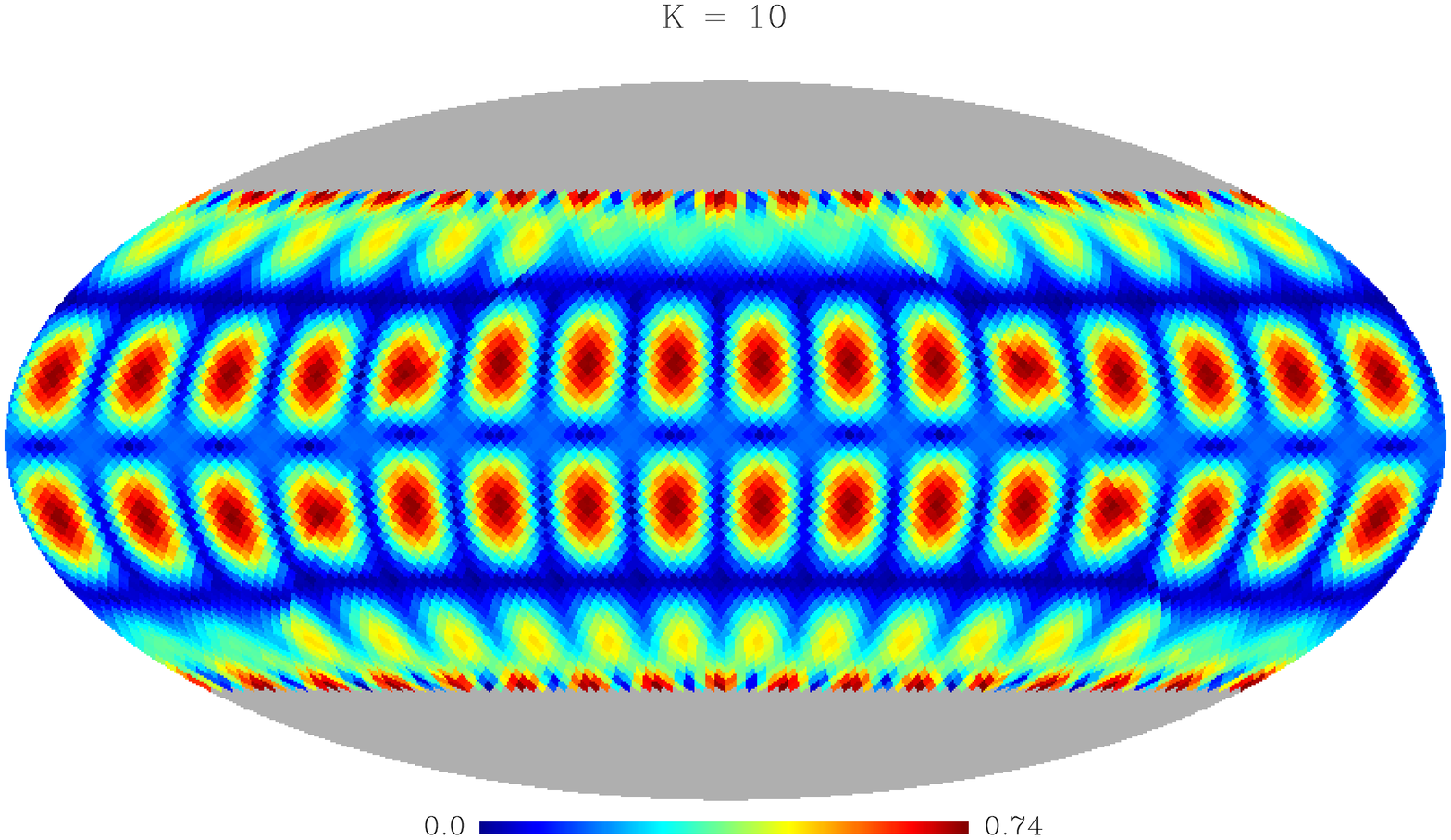}}\\
  \resizebox{0.6\hsize}{!}{\includegraphics[width=0.5\hsize,angle=90]{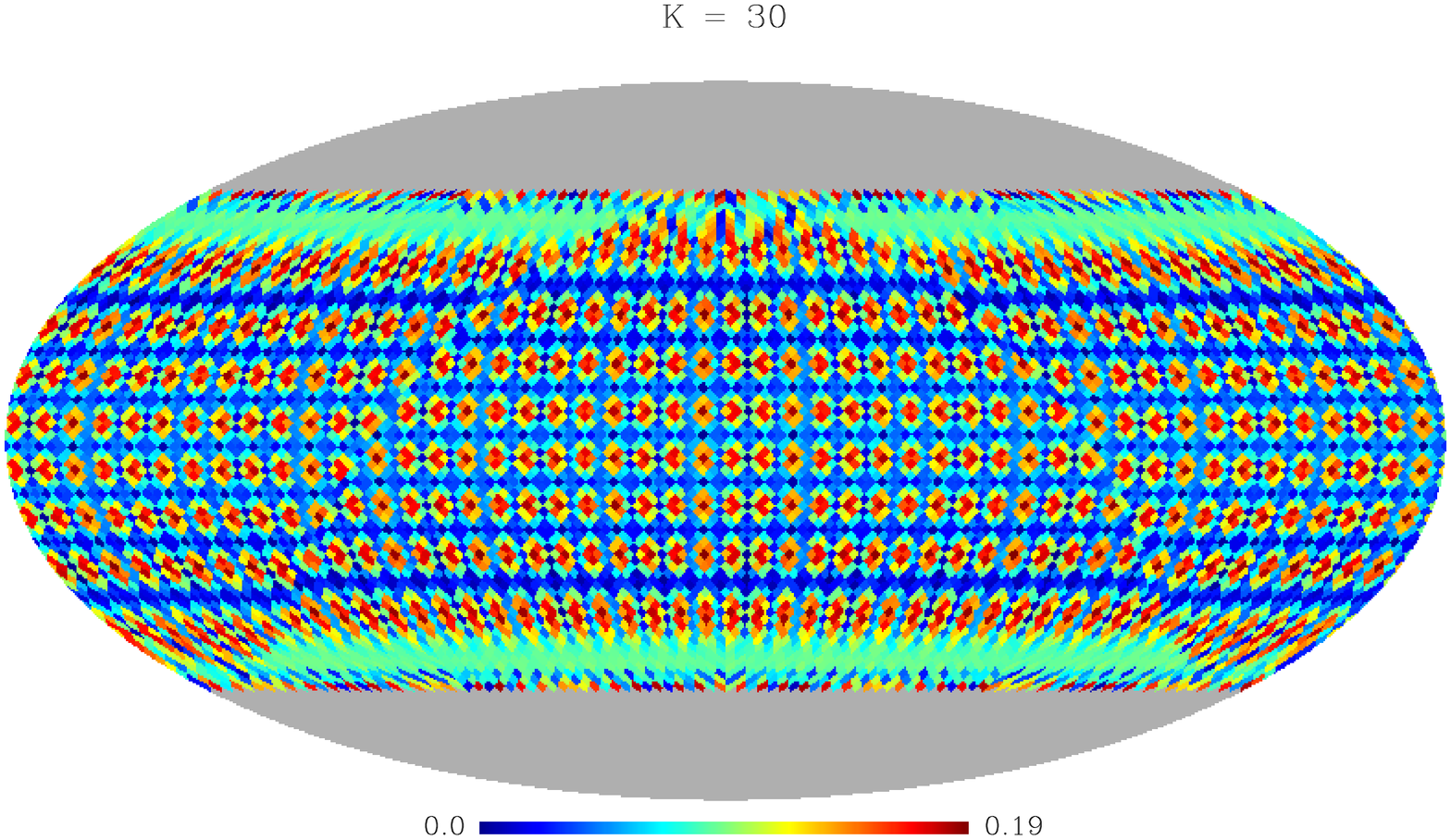}}
  \caption{
  Polarized intensity $I_p = \sqrt{Q^2 + U^2}$ map of the
  thermal contamination for a 12-month lifetime. The cases $K=1$ (top),
  $K=10$ (mid) and $K=30$ (bottom) are presented.
  All fluctuations are in arbitrary units considering
  $\left|\Delta(Q+jU)_0\right|=1$ the amplitude in one orbit.}
\label{contMapFig}
\end{figure*}
Clearly, the best case is the asynchronous one where the
fluctuations are well dumped on all the scales of interest for SPOrt
($\ell < 25$)
reaching values 3.5--5 orders of magnitude
lower than the instantaneous fluctuation amplitude  
($\sqrt{\ell(\ell+1)\;C_\ell/(2\pi)}$ is a fair
estimate of the signal).

The synchronous case with $K=1$ reduces the contamination as well,
but the dumping is limited to about 2.5 orders of magnitude.

The synchronous case with $K > 1$ is interesting:
dumping is relevant up to a critical $\ell_K$ at which the
fluctuations dramatically increase at a level much higher than
those for $K=1$. This confirms the results of the analytical
analysis (see Tab.~\ref{precDprTab}), where we find
that cases with $K>1$ can show maximum deviations larger than for
$K=1$.
But much more interesting is the relation between
the critical $\ell_K$ and
the fluctuation frequency: the higher $K$, the higher $\ell_K$.
That is, the high frequency of
the thermal fluctuations brings the power of the contamination
to small angular scales:
approximately, a $K=1$ behaviour generates dipole-like patterns, while, 
for a generic $K$, the dominant structures are approximately on 
$\theta = 180^\circ / K$ scale. 
This is confirmed by the spectral shapes of 
Fig.~\ref{spec12Fig} where the peak of the spectra
are at $\ell_{10} \sim 10$ for $K=10$ and at $\ell_{30} \sim 30$ for
$K=30$. 

The cleanliness of the $K=30$ configuration for $\ell < 25$ 
is appealing: 
all the $\ell$-range accessible to SPOrt has
low contamination 
(about 1-2 orders of magnitude lower than the case with $K = 1$),
making the condition $K>30$ an interesting option for the experiment.

The maps in Fig.~\ref{contMapFig}, reporting the polarized intensity
$I_p = \sqrt{Q^2 + U^2}$ of the contamination with 12-month lifetime,
support this view giving an insight from the pixel--space point of view. 
The patterns look different depending on the $K$ value and show
that the higher $K$, the smaller the size of the dominant structures. Thus,
a $K$ value large enough to make the dominant structures on scales
smaller than $7^\circ$ (SPOrt's FWHM) allows
the minimization of the thermal fluctuation impact in the angular-scale
range of interest for SPOrt.

It is worth noting that
the maximum values of the fluctuations in Fig.~\ref{contMapFig}
are close to those of Tab.~\ref{precDprTab}, 
in agreement with the analytical analysis.

The results for a longer lifetime (36~months) are shown in Fig.~\ref{spec12Fig}
as well, and their comparison with 12-month spectra gives us 
a hint to the time behaviour.
The case $K=1$ is practically unchanged, confirming
that a longer experiment does not provide a benefit in case of
synchronous fluctuations with $K=1$ (see Tab.~\ref{precDprTab}).

The cases $K>1$, instead, show a decrement of the thermal fluctuation
contamination, confirming that this configuration is
good for the SPOrt case, at least for $K>30$.
Finally, the asynchronous case does not show any improvement, in contrast
with the prediction of Tab.~\ref{precDprTab}. 
The numerical error limit is likely to have already been
reached.

\section{$E$-mode Contamination in the SPOrt case}\label{sportSec}

The SPOrt experiment has two well defined thermal environments 
(see also Tab.~\ref{phTempTab}): 
\begin{itemize}
  \item{} the horn location with temperature $T_{\rm ph}^{\rm h} \sim 300$~K.
          Simulations considering the ISS environment show that the natural 
          thermal fluctuations, with no thermal control, 
          are synchronous with the Sun illumination
          ($f_{\rm tf} = f_{\rm day}$) with amplitude $\Delta T \sim 3$~K. 
          However, the adoption of an active control performed through heaters 
          ensures a stability
          within $\Delta T_{\rm ph}^{\rm h} < 0.2$~K and
          breaks the correlation with the Sun modulation, moving the
          fluctuations to shorter periods with frequencies 
          $f_{\rm tf} > 30\;f_{\rm day}$ 
          for all the SPOrt radiometers. This
          occurs already for the 22~GHz, which,
          having the larger heat capacity, is the slowest. In addition, the
          fluctuations are decoupled from the Sun illumination modulation
          showing an asynchronous behaviour.
  \item{} the cryogenic part, including polarizer and OMT, 
          with temperature $T_{\rm ph}^{\rm cryo} \sim 80$~K
          and thermal stability within 
	        $\Delta T_{\rm ph}^{\rm cryo} < 0.1$~K 
	        provided by a closed-loop cryo-cooler. 
	        These fluctuations are intrinsically
	        uncorrelated with the Sun illumination.
\end{itemize}
\begin{table}
 \centering
  \caption{Temperature and fluctuations (maximum amplitude)
           of the SPOrt antenna devices.}
  \begin{tabular}{@{}lcc@{}}
      \hline
     Device     & $T_{\rm ph}$~[K] & $\Delta T_{\rm ph}$~[K]\\
     \hline
     Horn       & 300     &    $< 0.2$  \\
     Polarizer  & 80      &    $< 0.1$  \\
     OMT        & 80      &    $< 0.1$  \\
     \hline 
\end{tabular}
\label{phTempTab}
\end{table}
The presence of two thermal environments suggests to rewrite 
Eq.~(\ref{offFluct1eq}) as
\begin{eqnarray}
 \Delta(Q+jU)   & = &
           H_{\rm h}\; \Delta T_{\rm ph}^{\rm h}\nonumber \\
			 &+& (H_{\rm p}+H_{\rm omt})\;  \Delta T_{\rm ph}^{\rm cryo}
		  \label{sportOffFlucteq}
\end{eqnarray}
and to discuss their effects separately.

\subsection {Temperature Fluctuations of the Horn}

The offset fluctuations induced by the horn alone are
given by
\begin{eqnarray}
 \Delta(Q+jU) &=& {3\over 2}\,(A_{\rm h} - 1)\nonumber\\
     & &\times
     \;[S\!P_{\rm omt}(1+h^{\rm omt}) + S\!P_{\rm pol}(1+h^{\rm p})]
     \;\Delta T_{\rm ph}^{\rm h}.\nonumber \\
 \label{hornOffFlucteq}
\end{eqnarray}
Table~\ref{hornFluctTab} lists the values computed
for the SPOrt receivers at 22 and 90~GHz,
representing the best and worst cases, respectively.
\begin{table*}
 \centering
  \caption{Maximum offset fluctuation $\Delta(Q+jU)$ in one orbit (90~min)
           induced by the horn temperature instabilities
           for the SPOrt experiment. The cases of  the 22 and 90~GHz channels
           are reported, representing the best and worst cases among
           the SPOrt receivers. Details about horn temperature fluctuations 
	   ($\Delta T_{\rm ph}^{\rm h}$) and antenna characteristics are also
	   listed: horn attenuation ($A_{\rm h}$), 
	   OMT isolation ($|S_{B1}|^2$),
	   polarizer differential attenuation 
	   ($A_{\perp}-A_{\parallel}$), 
	   $S\!P_{\rm omt}$ and $S\!P_{\rm pol}$ coefficients, and
	   the $H_{\rm h}$ transfer function.
     $S\!P_{\rm omt}$ has been estimated with the approximation 
     $S\!P_{\rm omt} \sim 2\; A_{\rm omt}|S_{A1}S_{B1}^*|$ 
     as described in Carretti et al. (2001).
     The case of a 3~K variation is also shown, that is 
     the natural horn temperature fluctuation without active control.} 
  \begin{tabular}{@{}lccccccccc@{}}
     \hline
     Configuration     & $\nu$ & $\Delta T_{\rm ph}^{\rm h}$ & 
     $A_{\rm h}$  & $|S_{B1}|^2$  & 
     $A_{\perp}-A_{\parallel}$  & $S\!P_{\rm omt}$  &
     $S\!P_{\rm pol}$ & $H_{\rm h}$ & $\Delta(Q+jU)$  \\
             & [GHz] &  [K] & 
      [dB] &  [dB] & 
      [dB] &    &    &    &
      [$\mu$K]  \\
     \hline
     22GHz       & 22     &  $\pm \;0.2$  & 0.05 & -65 & -33 
      & $1.15\times 10^{-3}$  & $2.45\times 10^{-4}$ & $2.40\times 10^{-5}$ & 
     4.8\\
     90GHz       & 90     &  $\pm \;0.2$  & 0.1  & -60 & -30 
      & $2.05\times 10^{-3}$  & $4.89\times 10^{-4}$ & $8.75\times 10^{-5}$ &
     17.5\\
     22GHz (3~K) & 22     &  $\pm \;3$    & 0.05 & -65 & -33 
      & $1.15\times 10^{-3}$  & $2.45\times 10^{-4}$ & $2.40\times 10^{-5}$ & 
     72.0\\
     90GHz (3 K) & 90     &  $\pm \;3$    & 0.1  & -60 & -30 
      & $2.05\times 10^{-3}$  & $4.89\times 10^{-4}$ & $8.75\times 10^{-5}$ &
     262.0\\
     \hline\\
\end{tabular}
\label{hornFluctTab}
\end{table*}

A low sensitivity is clearly shown by the
transfer functions, whose levels reduce the impact of 
thermal instabilities by 
4--5 orders of magnitude. 
The offset fluctuations are directly related to the values of
the OMT isolation and the differential attenuation between the
two main polarizations of the polarizer 
(see Eqs.~(\ref{spomtEq}) and~(\ref{sppolEq})). 
Such low offset fluctuations are due to 
the improvements in performances of passive devices
obtained by the SPOrt team (Cortiglioni et al. 2004, Peverini et al. 2003). 
In fact, state-of-the-art OMTs available  
at the beginning of the project
had isolations worse than 40~dB, 
while the device developed for this experiment, with
about 60~dB isolation, leads to a decrease of the offset fluctuations
by a factor about 10.

As described in Sec.~\ref{statSec}, the evaluation of the thermal
instability impact has to be performed in the
multipole space, where the contamination
on different angular scales can be estimated and compared to the 
expected cosmological signal.

\begin{figure}
\centering
\resizebox{0.9\hsize}{!}{\includegraphics{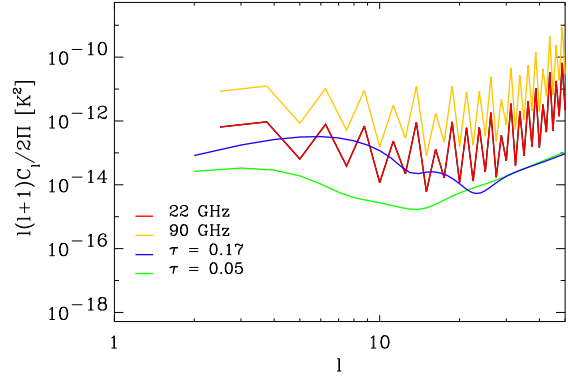}}
\caption{$E$--mode power spectra of the contamination induced by
thermal fluctuations of the horn 
in case of no temperature control. 
Spectra are corrected for the smearing effects of the beam window function.
A lifetime of 36 months is considered.
The $E$-mode spectrum expected
for the cosmological signal is also reported for comparison for two
cosmological models: the concordance model as from WMAP's
first-year results with $\tau = 0.17$ (Spergel et al. 2003);
a similar model but with a lower $\tau = 0.05$.
}
\label{specSP3KFig}
\end{figure}

First of all we consider what happens in the case of no
active control, corresponding to a synchronous behaviour 
with $K=1$.
The contamination-map power spectra
are shown in Fig.~\ref{specSP3KFig}, along with the
expected polarized sky emission ($E$-mode)
as from WMAP best-fit cosmological model
(Spergel et al. 2003, Kogut et al. 2003).
The cosmological signal appears significantly contaminated
even when the optical depth of the reionized medium is $\tau = 0.17$.
This means that the free thermal fluctuations induced by the ISS environment
are too large for such a measurement,
calling for a reduction of the thermal disturbances.
\begin{figure}
\centering
\resizebox{0.9\hsize}{!}{\includegraphics{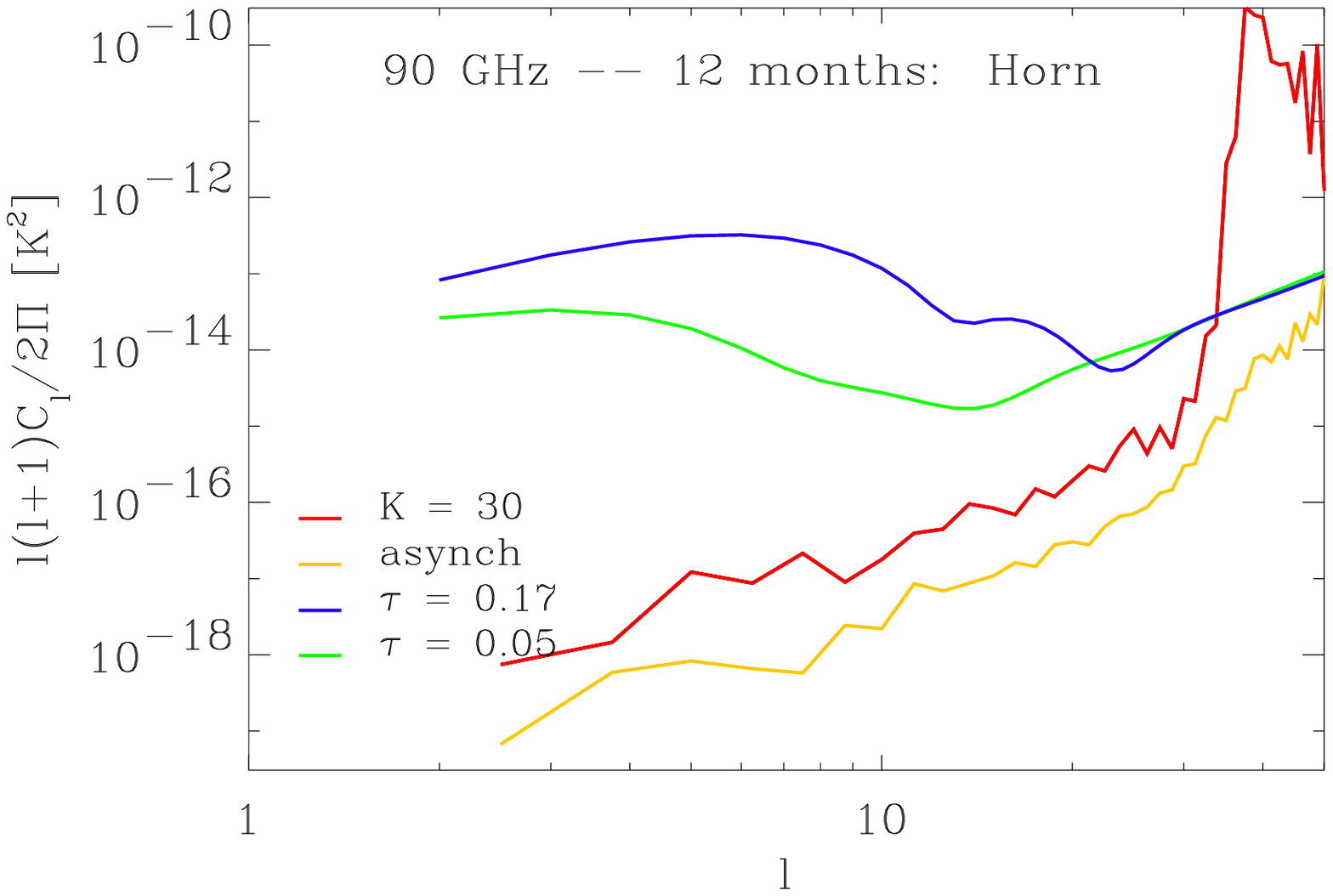}}
\resizebox{0.9\hsize}{!}{\includegraphics{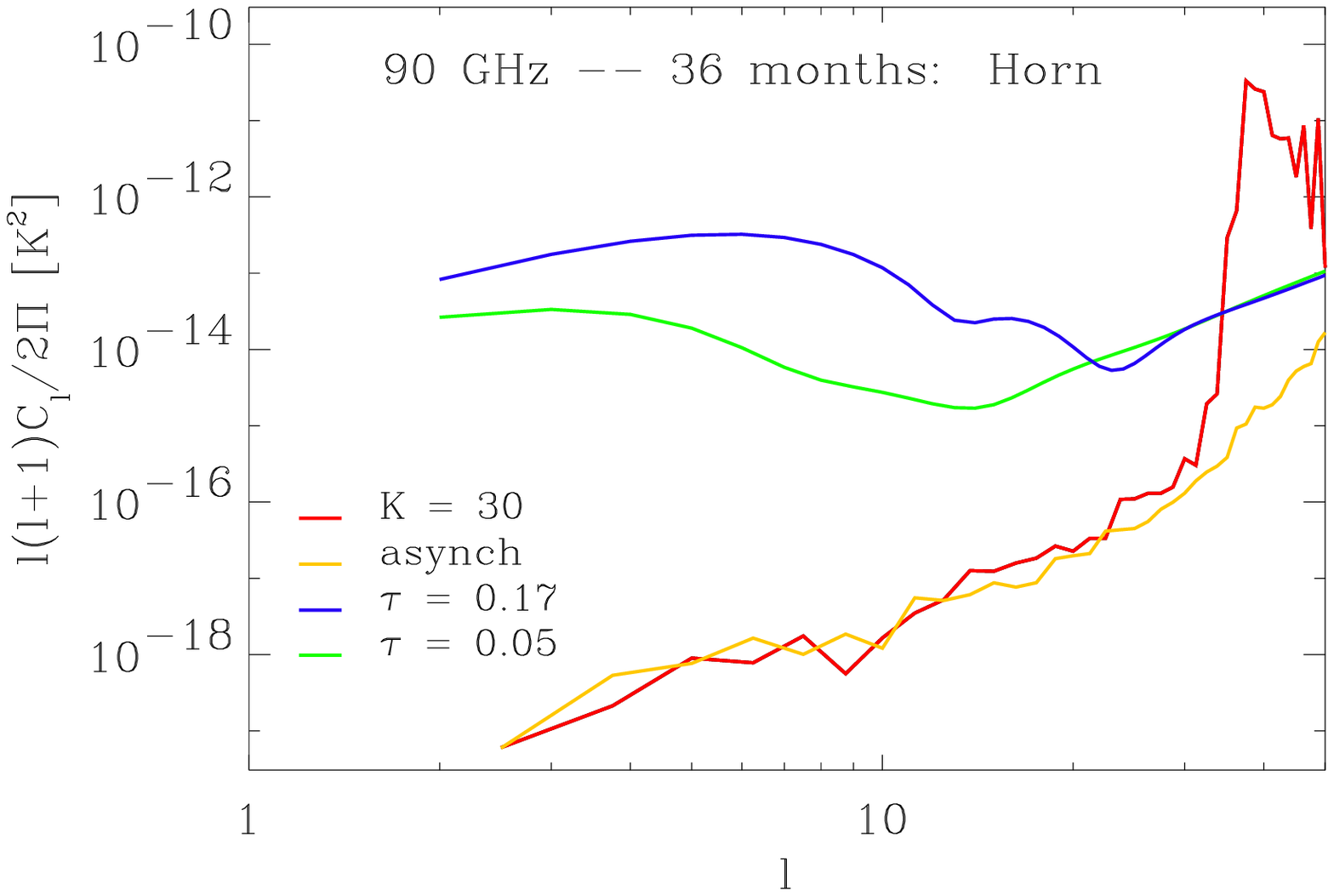}}
\caption{Top: $E$--mode power spectra of the 
contamination due to horn thermal fluctuations 
with active temperature control.
Spectra are corrected for the smearing effects of the beam window function.
The 90~GHz receiver and a lifetime of 12~months
are assumed.
The $E$-mode spectrum expected
for the cosmological signal is also reported for comparison for two
cosmological models: that from WMAP
first-year results with $\tau = 0.17$ (Spergel et al. 2003);
a similar model but with a lower $\tau = 0.05$.
Bottom: the same for a 36-month lifetime.}
\label{specSP12Fig}
\end{figure}

The active control adopted for SPOrt 
allows temperature fluctuations with
amplitude $\Delta T^{\rm h}_{\rm ph} < 0.2$~K and 
with frequency $f_{\rm tf} > 30 f_{\rm day}$,
which satisfies the condition identified 
in Sec.~\ref{statSec}. 
The power spectrum of the contamination for this configuration
is reported in 
Fig.~\ref{specSP12Fig}, where the worst case
(90~GHz receiver) is shown
for two different lifetimes (12 and 36 months).
Here we consider both
a synchronous behaviour with $K=30$ 
and the case of asynchronous fluctuations.
The main result is that already for $K=30$ and a 12-month lifetime
the contamination is well below
the signal, not only for $\tau = 0.17$, but even for 
a lower optical depth $\tau = 0.05$. 
A 36--month lifetime provides even better 
results. However, the 12-month lifetime already provides a
very comfortable scenario which does not need any specific 
data cleaning. 

\subsection {Temperature Fluctuations of the Cryo-Stage}

The cryo-stage induces offset fluctuations according to the formula
\begin{equation}
 \Delta(Q+jU) = (H_{\rm p} + H_{\rm omt})\; \Delta T_{\rm ph}^{\rm cryo}.
			    \label{tfColdeq}
\end{equation}
The values for the 22 and 90~GHz receivers are reported
in Tab.~\ref{cryoFluctTab}.
\begin{table*}
\centering
  \caption{As for Tab.~\ref{hornFluctTab}, but for the cryo-stage of the SPOrt radiometers.
  As a worst case, we assume 
  $A_{\rm p} = 0.1$~dB and $A_{\rm omt} = 0.2$~dB the attenuations of
  polarizer and OMT, respectively. Anyway, the offset fluctuations are
  marginally dependent on their values, $S\!P_{\rm pol}$ being
  the dominant term.} 
  \begin{tabular}{@{}lcccccccccc@{}}
     \hline
     Configuration     & $\Delta T_{\rm ph}^{\rm cryo}$ & 
     $A_{\rm h}$  & $|S_{B1}|^2$  & 
     $A_{\perp}-A_{\parallel}$  & $S\!P_{\rm omt}$  &
     $S\!P_{\rm pol}$ & $H_{\rm p}$ & $H_{\rm omt}$ & 
     $\Delta(Q+jU)$  \\
             &  [K] & 
      [dB] &  [dB] & 
      [dB] &    &    &    &    &
      [$\mu$K]  \\
     \hline
     22GHz       &  $\pm 0.1$  & 0.05 & -65 & -33 
      & $1.15\times 10^{-3}$  & $2.45\times 10^{-4}$ & 
      $3.21\times 10^{-4}$ & $8.62\times 10^{-5}$ & 
     40.7\\
     90GHz       &  $\pm 0.1$  & 0.1 & -60 & -30 
      & $2.05\times 10^{-3}$  & $4.89\times 10^{-4}$ & 
      $6.46\times 10^{-4}$ & $1.56\times 10^{-4}$ &
     80.2\\
     \hline\\
\end{tabular}
\label{cryoFluctTab}
\end{table*}
\begin{figure}
\centering
  \resizebox{0.9\hsize}{!}{\includegraphics{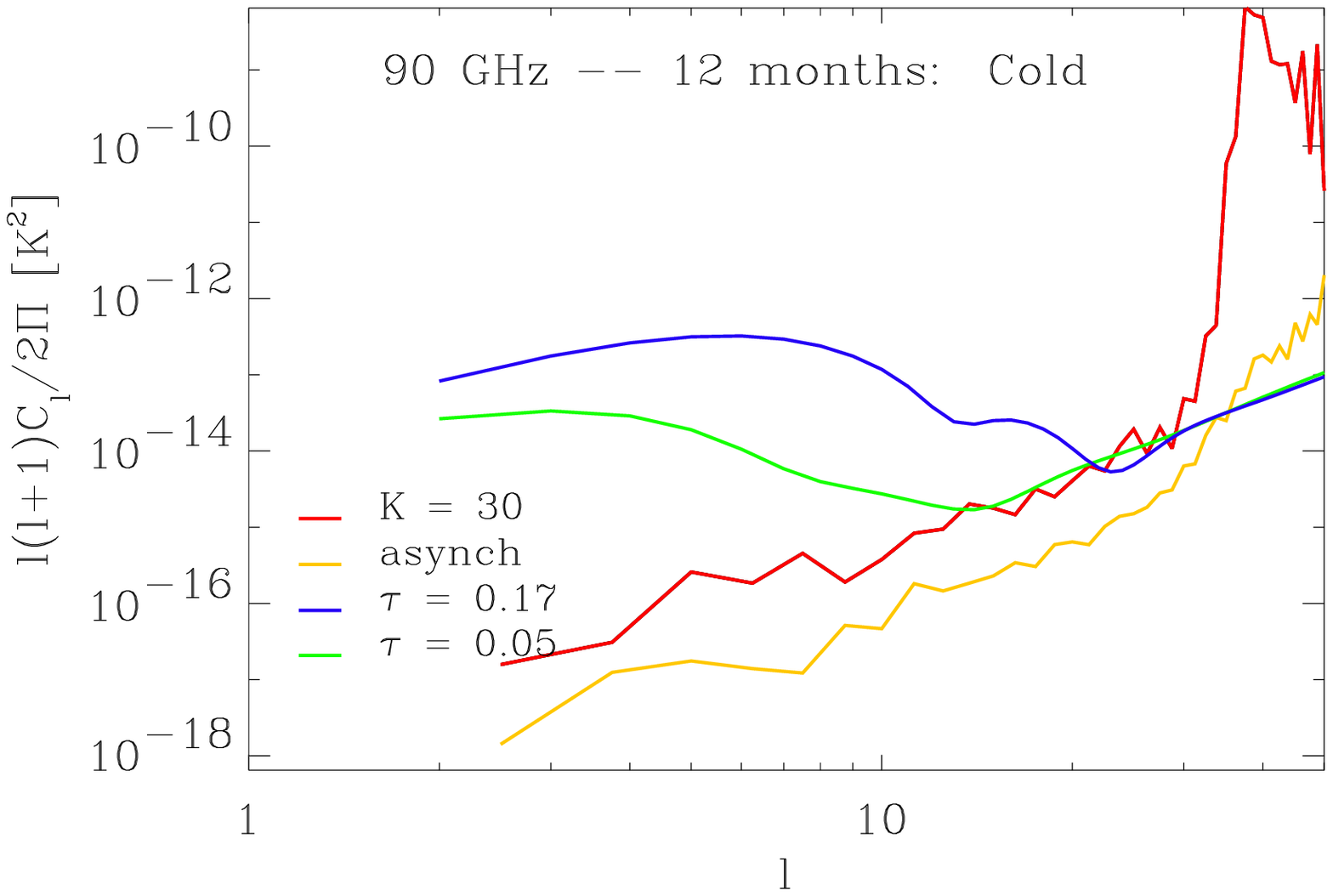}}
  \resizebox{0.9\hsize}{!}{\includegraphics{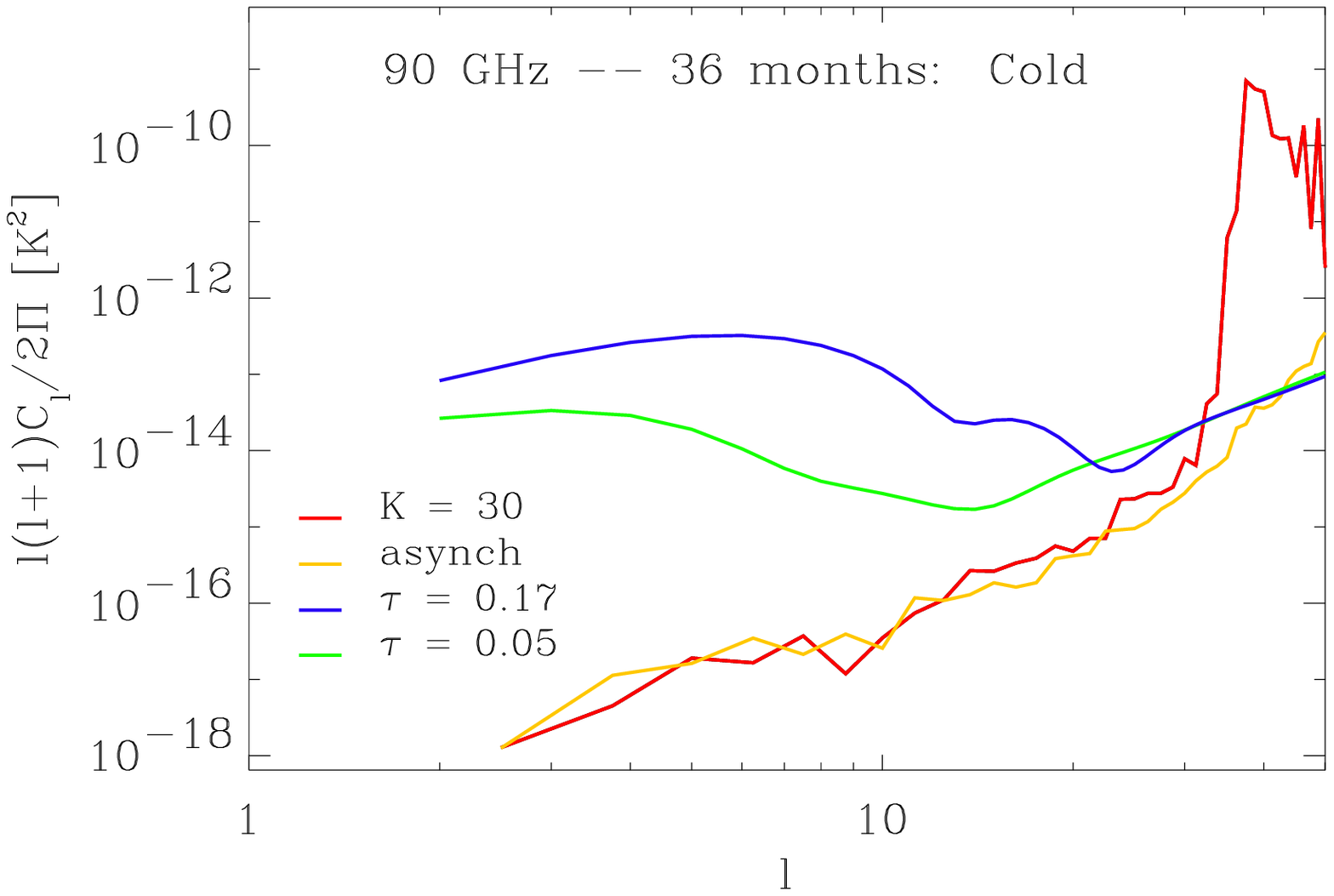}}
  \caption{As Fig.~\ref{specSP12Fig} but for the cryo-stage of the 
  90~GHz receiver.}
\label{specSPC12Fig}
\end{figure}

In this case the dominant term is related to the polarizer and
only $S\!P_{\rm pol}$ acts to dump the fluctuation, without
contributions from ($A-1$) terms. As already mentioned
by Carretti et al. (2001), the leading offset due to the
polarizer is a polarized noise generated by the device itself,
rather than the correlation of an unpolarized noise generated by the antenna.
This is why only one dumping factor is in action ($S\!P_{\rm pol}$), 
leading to the quite unexpected result that
the cryo-stage generates the most
relevant contamination, larger than the horn contribution,
even though the latter is in a warm section. 

The behaviour is determined by the cooler electronics, which, performing
an active control to anchor the temperature to a fixed set-point, induces
fluctuations not related with the external environment. 
This is true only as a first approximation, since
a small correlation can arise due to the coupling between the
cryo-stage and the feed horn. In fact, these are connected through a 
transition waveguide whose aim is to thermally separate the two 
environments while allowing the passage of electromagnetic waves.
However, the thermal separator filters the fluctuations
and only a small fraction is transmitted to the cold-stage.
As an example, tests performed during the integration phases 
on the BaR-SPOrt receiver,
an instrument similar to SPOrt (Cortiglioni et al. 2003),
show that fluctuations of the cold-stage induced by the horn 
are a few percent of those of the horn itself. Therefore,
considering a 0.2~K horn temperature variation, 
the fluctuations correlated to the warm part are
expected to be no more than few hundredths of Kelvin.
Anyway, should a coupling with the horn arise, 
the fluctuations would have a statistics similar to that of the
horn, leading to, as the worst case, a periodic behaviour with
$f_{\rm tf} > 30\;f_{\rm day}$.

As for the horn section, we study the
contamination impact through the analysis of the power spectrum.
We perform the analysis for two configurations: the 
asynchronous behaviour and the synchronous one with $K=30$, the latter
representing the worst condition, especially if all the 0.1~K variation
is considered of such a kind. 
The spectra are reported in  Fig.~\ref{specSPC12Fig} 
for two different mission durations.
Although higher than that 
from the horn section, the contamination is again
well below the sky signal already for a 12-month
mission and synchronous fluctuations with $K=30$.
Also this contribution, thus, is unlikely to require any
data cleaning.
It is worth noting that even the low $\tau = 0.05$ model
is free from contaminations in the $\ell$-range
where most of the cosmological signal resides ($\ell < 10$).

\section{Summary and Conclusions}\label{concSec}

In this work the importance of the errors induced by thermal instabilities
has been evaluated for the SPOrt experiment. 
In particular we have computed and analyzed
\begin{enumerate}
  \item the transfer functions of the front-end passive devices (horn, OMT, polarizer)
        describing how the temperature fluctuations are reflected on the signal;
  \item the dumping effects on the final sky maps due to both the
        SPOrt scanning strategy and the behaviour of the fluctuations.
\end{enumerate}
The main results can be summarized as follows:
\begin{itemize}
  \item The correlation scheme of SPOrt has a low sensitivity
        to temperature instabilities. In fact, the transfer functions
        are the products between ($A-1$) terms, common to total power
        architectures, and $S\!P_{\rm pol}$ and $S\!P_{\rm omt}$ coefficients,
        specific to correlators. Thus, these benefit from extra-terms
        which further reduce the impact of thermal fluctuations,
        making correlators a suitable solution for measurements
        of weak signals;
  \item The 70-day precession of the ISS orbit allows cancellation
        effects in the case of fluctuations synchronous with
        the Sun illumination ($K=1$). The dumping is more than 2 orders of magnitude.
  \item Fluctuations with shorter periods ($K > 1$) show stronger
        dumping, but only on angular scales larger than $180^\circ/K$. In fact, 
        they move the contamination to multipoles $\ell > \ell_K \sim K$, leaving
        the range $\ell < \ell_K$ much cleaner than in the
        case with $K=1$. As a consequence,
        the condition $K>30$ is enough to minimize the contamination on 
        the angular scales accessible to SPOrt ($\ell < 25$).
  \item Fluctuations uncorrelated with the Sun illumination represent
        the best condition, although their contamination level is not so far
        from the synchronous case with $K>30$;
\end{itemize}
We would like to point out
that the results for $K=1$ are mainly due to different Sun
illumination conditions during the different precession periods. 
Therefore, in case of
no changes in the thermal conditions during different scans, 
this dumping effect would not be present.
An example of such a condition is the Lagrangian point L2
of the Earth-Sun system, where the same 
pixel is observed every 6~months with similar
illumination conditions (Sun in the back
of the spacecraft). L2 enjoys an optimal environment stability, 
much better than that in the low-orbit of ISS.
However, no dumping 
effects due to different Sun illumination conditions can be expected, so that
the impact of scan synchronous fluctuations must
be carefully evaluated,
especially when dealing with the weak CMBP signal. In this case relevant benefits
can only be obtained from a proper scanning strategy with several crossing 
at different angles among the scans, as that of WMAP.

The analysis of the actual antenna system of SPOrt shows a low sensitivity to
thermal instabilities, achieved thanks to both the correlation scheme and 
new devices developed by the SPOrt team to minimize systematic effects.
In spite of this, the fluctuations induced 
by the ISS orbit would be too large to allow a clean detection of CMBP
in absence of an active thermal control of the horns. 
On the other hand, the active control adopted for the horns
leads to fluctuations featuring an amplitude within $\pm0.2$~K (instead of 
$\pm3$~K) and a frequency $f_{\rm tf} > 30\; f_{\rm day}$.
Our analysis shows that the resulting contamination is well
below the expected cosmological signal, leaving
the $E$-mode spectrum of the CMBP uncontaminated.

The thermal fluctuations of the cold-stage
generate a contamination low enough to allow again 
a clean detection of
the CMBP signal, although this is - surprisingly - 
the most contaminant source, even larger than that of the warm stage. 
 
The cold stage contribution, being much larger than that from the warm section,
in practice represents the total contamination for the SPOrt experiment,
which thus, in spite of an unfriendly environment, appears to be robust 
against systematics induced by thermal fluctuations.


\section*{Acknowledgments}
This work has been carried out in the frame of the SPOrt experiment, 
a programme of the Italian Space Agency
(Agenzia Spaziale Italiana: ASI). 
We thank Riccardo Tascone for useful discussions
and the referee for useful comments.
MZ, CM and CS acknowledge ASI grants. 
Some results of this paper have been derived using the 
HEALPix\footnote{http://www.eso.org/science/healpix/} (G\'orski, Hivon
\& Wandelt 1999). We acknowledge the use of the CMBFAST package.

\appendix

\section{Temperature Dependence of Attenuation, $S\!P_{\rm pol}$
          and $S\!P_{\rm omt}$}\label{AppA}

In general, the attenuation $A$ of passive devices like feed horns, polarizers and OMTs
dependes on the electric resistivity in a fashion (e.g. see Collin 1996)
\begin{equation}
  A_{\rm dB} \propto \sqrt{\rho}.
\end{equation}
where $A_{\rm dB}$ is expressed in dB unit.
In the temperature range of interest for SPOrt (80-300~K), the resistivity $\rho$
has a linear behaviour with respect to the temperature $T$ for
most of the materials the devices are made of (e.g. Al, Au, Ag), so that
\begin{eqnarray}
  A_{\rm dB} &\propto& \sqrt{T}
\end{eqnarray}
that means the linear attenuation follows
\begin{eqnarray}
  A &=& {\rm e}^{K_A\sqrt{T}}
\end{eqnarray}
where $K_A$ is a proper constant.
At lower temperature, the linear behaviour of $\rho$
is no more valid and a proper function of $T$ has to be used.
However, this dependence weekens ($\rho$
is going to be constant toward $T$~=~0~K) and the variations we derive below
can be considered as a worst case estimate.

For devices with low attenuation ($A = 1+x$, with $x$~$<$$<$~1) a linear approximation
can be applied. In this limit the simple relation
\begin{eqnarray}
  A       &\sim& 1 + K_A \;\sqrt{T}
\end{eqnarray}
holds and
the variations with respect to $T$ of the attenuations and
the $S\!P_{\rm pol}$, $S\!P_{\rm omt}$ coefficients
can be easily computed. In particular we have the following
results:
\begin{enumerate}
\item Attenuations: Expanding to the first order of $K_A \,\sqrt{T}$ one
      gets
      \begin{eqnarray}
          \Delta A &=& {K_A \,\sqrt{T}\over 2\,T} \; A \;\Delta T\nonumber\\
                   &\sim& {K_A \,\sqrt{T}\over 2\,T} \;\Delta T\nonumber\\
                   &\sim& {A-1\over 2\,T}\;\Delta T.
      \end{eqnarray}
      Alternatively, starting from the first row of the previous equation,
      one can also write
      \begin{eqnarray}
          \Delta A &\sim& {A-1\over 2\,T}\;A\;\Delta T.
      \end{eqnarray}
      In this paper we use both of them.
      
\item $A-1$: In similar way, it can be written
      \begin{eqnarray}
          \Delta (A-1) &\sim& {A-1\over 2\,T}\;\Delta T,
      \end{eqnarray}
      or, equivalently, 
      \begin{eqnarray}
          \Delta (A-1) &\sim& {A-1\over 2\,T}\;A\;\Delta T.
      \end{eqnarray}
\item $S\!P_{\rm pol}$ coefficient: Eq.~(\ref{sppolEq}) can be written as 
      \begin{eqnarray}
          S\!P_{\rm pol} & = & {1\over 2} \left(1 - {A_{\parallel}\over
                                 A_{\perp}}\right) \nonumber\\
                         & = & {1\over 2} 
                               \left(1-{\rm e}^{(K_{\perp}-K_{\parallel})\;\sqrt{T}}
                               \right)\nonumber\\
                         &\sim& {1\over 2} 
                               (K_{\perp}-K_{\parallel})\;\sqrt{T}
      \end{eqnarray}
      where $K_{\perp}$ and $K_{\parallel}$ are the $K_A$ coefficients of
      $A_{\perp}$ and $A_{\parallel}$, respectively. Therefore,
      the variation of $S\!P_{\rm pol}$ writes
      \begin{eqnarray}
          \Delta S\!P_{\rm pol} & \sim &  {1\over 2} 
                               (K_{\perp}-K_{\parallel})\;\sqrt{T}
                               \; {\Delta T\over 2\,T} \nonumber\\
                               & \sim &  {S\!P_{\rm pol}\over 2\,T} \; \Delta T
      \end{eqnarray}
\item $S\!P_{\rm omt}$ coefficient: This contains not only attenuation terms, 
      but also the isolations $S_{A2}$, $S_{B1}$ between the two OMT arms.
      The latter mainly depends on the geometry of the device rather than
      on resistive effects, so that, as a first approximation, 
      they can be considered constant with respect
      to the temperature. Eq.~(\ref{spomtEq}) can be written as 
      \begin{eqnarray}
           S\!P_{\rm omt} & = & S\!P_{\rm omt}^A + S\!P_{\rm omt}^B,
      \end{eqnarray}
      with 
      \begin{eqnarray}
           S\!P_{\rm omt}^A & = & A_{\rm omt}\;S_{A1}S_{B1}^*,\nonumber\\
           S\!P_{\rm omt}^B & = & A_{\rm omt}\;S_{A2}S_{B2}^*.
      \end{eqnarray}
      Reminding that the attenuations along the OMT arms are given by 
      $A_{\rm omt}^A = 1/|S_{A1}|^2$, $A_{\rm omt}^B = 1/|S_{B2}|^2$, one can
      write
      \begin{eqnarray}
           \Delta S\!P_{\rm omt}^A & \sim & {A_{\rm omt}-1 \over 2\,T} 
                                          \; A_{\rm omt}\;S_{A1}S_{B1}^* \; \Delta T\nonumber\\
                                   & &      - {A_{\rm omt}^A-1 \over 4\,T} 
                                          \; A_{\rm omt} \;S_{A1}S_{B1}^*\; \Delta T\nonumber\\
                                   & = &    \left[ (A_{\rm omt}-1) -
                                            {1\over 2}\; (A_{\rm omt}^A-1)\right]\; 
                                            {S\!P_{\rm omt}^A \over 2\,T}\;\Delta T.
                                            \nonumber\\
      \end{eqnarray}
      The OMT attenuation $A_{\rm omt}$ is the average of those along the two arms, 
      for which, in general, the relation $A_{\rm omt}^A \sim A_{\rm omt}^B 
      \sim A_{\rm omt}$ holds. Thus, we can write
      \begin{eqnarray}
           \Delta S\!P_{\rm omt}^A & \sim & {A_{\rm omt}-1 \over 4\,T}\;
                                            S\!P_{\rm omt}^A\; \Delta T.
      \end{eqnarray}
      A similar relation holds for the other term       
      \begin{eqnarray}
           \Delta S\!P_{\rm omt}^B & \sim & {A_{\rm omt}-1 \over 4\,T}\;
                                            S\!P_{\rm omt}^B\; \Delta T,
      \end{eqnarray}
      so that the variation of $S\!P_{\rm omt}$ is given by
      \begin{eqnarray}
           \Delta S\!P_{\rm omt} & = & \Delta S\!P_{\rm omt}^A + \Delta S\!P_{\rm omt}^B 
                                     \nonumber \\
                                 & \sim & {A_{\rm omt}-1 \over 4\,T}\;
                                            S\!P_{\rm omt}\; \Delta T.
      \end{eqnarray}
\end{enumerate}

\end{document}